\documentclass[twocolumn,showpacs,prb,10pt]{revtex4-1}
\usepackage[english]{babel}
\usepackage{amsmath,amssymb} 
\usepackage{times}
\usepackage{epsfig} 
\usepackage{placeins} 

\newcommand{\eq}[1]{\begin{equation}#1\end{equation}}
\newcommand{\eqa}[1]{\begin{eqnarray}#1\end{eqnarray}}
\newcommand{\tr}{\textrm{Tr}}

\newcommand{\aver}[1]{\langle#1\rangle}

\usepackage[colorlinks=true, pdfstartview=FitV, linkcolor=blue,
citecolor=blue, urlcolor=blue]{hyperref} 
\usepackage{hyperref}

\usepackage{color}
\definecolor{nred}   {RGB}{224,0,0}

\begin{document} 

\title{Enhancement of the
thermal expansion of organic charge transfer salts\\
by strong electronic correlations
}

\author{J. Kokalj$^1$}
\email{jure.kokalj@ijs.si}
\author{Ross H. McKenzie$^2$}
\email{r.mckenzie@uq.edu.au}
\homepage{condensedconcepts.blogspot.com}
\affiliation{$^1$J.\ Stefan Institute, SI-1000 Ljubljana, Slovenia}
\affiliation{$^2$School of Mathematics and Physics, University of Queensland,
  Brisbane, 4072 Queensland, Australia} 
\date{\today}

\begin{abstract}
  Organic charge transfer salts exhibit thermal expansion anomalies
  similar to those found in other strongly correlated electron
  systems. The thermal expansion can be anisotropic and have a
  non-monotonic temperature dependence. We show how these anomalies
  can arise from electronic effects and be significantly enhanced,
  particularly at temperatures below 100 K, by strong electronic
  correlations.  For the relevant Hubbard model the thermal expansion
  is related to the dependence of the entropy on the parameters ($t$,
  $t'$, and $U$) in the Hamiltonian or the temperature dependence of
  bond orders and double occupancy.  The latter are calculated on
  finite lattices 
  with the Finite Temperature Lanczos Method.  Although many features
  seen in experimental data, in both the metallic and Mott insulating
  phase, are described qualitatively, the calculated magnitude of the
  thermal expansion is smaller than that observed experimentally.
\end{abstract}

\maketitle 

\section{Introduction}

The thermal expansion coefficients of a wide range of strongly
correlated electron materials exhibit temperature and orientational
dependencies that are distinctly different from simple metals and
insulators\cite{white1993}.  Materials that have been studied included
heavy fermion compounds \cite{lacerda89}, organic charge transfer
salts \cite{muller02,souza07,manna10,desouza}, iron pnictide
superconductors \cite{meingast2012,hardy13}, and LiV$_2$O$_4$
\cite{johnston}.  The Gr\"uneisen parameter $\Gamma$, which is
proportional to the ratio of the thermal expansion to the specific
heat, can be two orders of magnitude larger than the values of order
unity found for elemental solids \cite{lacerda89}, and may diverge at
a quantum critical point \cite{zhu2003}.  For organic charge transfer
salts, the thermal expansion coefficients show anomalies at the
superconducting transition temperature \cite{muller00}, at the Fermi
liquid coherence temperature, at the Mott transition, and strong
non-monotonic temperature and orientational dependence
\cite{muller02}.  Anomalies have been recently observed also in a spin
liquid candidate material, $\kappa$-(BEDT-TTF)$_2$Cu$_2$(CN)$_3$ \cite{manna10}. 
 For a proper understanding
and interpretation of these experimental results it is important to
elucidate the electronic (apart from the phononic) contribution to the
thermal expansion, particularly since it may dominate at low
temperatures.  Related electronic effects are seen in lattice
softening near the Mott transition via sound velocity measurements
\cite{fournier03,hassan05}.  The electronic contribution can also lead
to the critical behaviour of the thermal expansion close to the
metal-insulator transition \cite{souza07,zacharias12}.

Here we study the electronic contribution to the thermal expansion,
including its directional dependence, by modelling the electrons with
a Hubbard model on the anisotropic triangular lattice at half filling,
an effective model Hamiltonian for several families of organic charge
transfer salts \cite{powell11}. Our analysis requires a connection
between the Hubbard model parameters ($t$, $t'$, $U$) and structural
parameters (lattice constants) that can be deduced from electronic
structure calculations and bulk compressibilities for which we use
experimental values.

\subsection{Summary of results}

Our main results concerning the electronic contribution to the thermal
expansion $\alpha$ are as follows.

(i) At low temperatures strong correlations can increase the thermal
expansion by as much as an order of magnitude.

(ii) A non-monotonic temperature dependence of $\alpha$ is possible.

(iii) Significant orientational dependence is possible, including the
expansion having the opposite sign in different directions.

(iv) In the metallic phase the crossover from a Fermi liquid to a bad
metal may be reflected in a maximum in the temperature dependence of
$\alpha$.

(v) In the Mott insulating phase a maximum in the temperature
dependence of $\alpha$ can occur, at a temperature comparable to that
at which a maximum also occurs in the specific heat and the magnetic
susceptibility.

(vi) All of the above results 
are sensitive to the proximity to the Mott metal-insulator
transition and the amount of frustration, reflected in the parameter
values ($U/t$ and $t'/t$) in the Hubbard model.

Although, we can describe many of the unusual qualitative features of
experimental data for organic charge transfer salts, the overall
magnitude of the thermal expansion coefficients that we calculate are
up to an order of magnitude smaller than observed. This disagreement
may arise from uncertainties in how uniaxial stress changes the
Hubbard model parameters, and uncertainty in the compressibilities
including not taking into account the effect of softening of the
lattice associated with proximity to the Mott transition.

\subsection{Specific experimental results we focus on}

\begin{figure}[htb] 
 \centering 
\includegraphics[width =42mm,angle=-90]{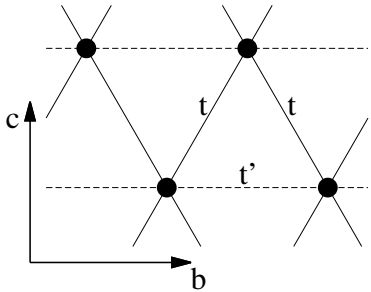}
  \caption{Lattice and hopping integrals in the Hubbard model for
$\kappa$ -(BEDT-TTF)$_2$X
with X=Cu$_2$(CN)$_3$ and Cu(NCS)$_2$.
For X= Cu[N(CN)$_2$]Br the crystal axes
$b$ and $c$ should be replaced with
$c$ and $a$. 
 }
\label{fig_0}
\end{figure}

We briefly review some experimental results that our
calculations are directly relevant to.
We only consider thermal expansion within the conducting layers.
Anomalies are also seen in the interlayer direction but are beyond the
scope of this study.  Figure \ref{fig_0} shows the relation between
the anisotropic triangular lattice, and the associated hoppings $t$
and $t'$, and the intralayer crystal axes ($b$ and $c$) for $\kappa$
-(BEDT-TTF)$_2$X with anions X=Cu$_2$(CN)$_3$ and Cu(NCS)$_2$.  For X=
Cu[N(CN)$_2$]Br, the crystal axes $b$ and $c$ should be replaced with $c$ and $a$,
respectively. 

Mott insulating phase of $\kappa$-(BEDT-TTF)$_2$Cu$_2$(CN)$_3$
[Figure 1 in Ref. \onlinecite{manna10}].  For some temperature ranges
thermal expansion in the $b$ ($t'$) and $c$ ($t$) directions have
opposite signs.  Thermal expansion is a non-monotonic function of
temperature.  $\alpha_b$ and $\alpha_c$ have extremal values at about
60 K and 30 K, respectively. For comparison, the magnetic
susceptibility has a maximum at a temperature around 60 K
\cite{shimizu03}.

Metallic phase of un-deuterated
$\kappa$-(BEDT-TTF)$_2$Cu[N(CN)$_2$]Br  [Figure 5a in
Ref. \onlinecite{muller02}].  As the temperature decreases there is a
crossover from a bad metal to a Fermi liquid.\cite{merino2008}
$\alpha_a$ is a 
non-monotonic function of temperature, with a large maximum around 35
K, which is comparable to the temperature at which there is a large
change in slope of the resistivity versus temperature curve
\cite{yu91}. This is one measure of the coherence temperature
associated with the crossover from the bad metal to the Fermi
liquid.\cite{deng12}
 
Mott transition in fully deuterated
$\kappa$-(BEDT-TTF)$_2$Cu[N(CN)$_2$]Br  [Figure 1 in
Ref. \onlinecite{souza07} and Figure 2 in Ref. \onlinecite{lang2008}].
As the temperature decreases there is a crossover from a bad metal to
a Fermi liquid to a Mott insulator (below 14 K).  $\alpha_c$
(direction of $t'$) is much smaller than $\alpha_a$ (direction of $t$)
and monotonically increases with temperature.  In contrast,
$\alpha_a$ is a non-monotonic function of temperature, with a large
maximum around 30 K, which is comparable to the temperature at which
the crossover from the bad metal to the Fermi liquid occurs.  Also,
$\alpha_a$ is negative in the Mott insulating phase.

\subsection{Outline}

The outline of the paper is as follows.  In Section \ref{sec:thermo}
we discuss how the thermal expansion is related to variations in the
entropy through Maxwell relations from
thermodynamics.  In Section \ref{sec:hubbard} the relevant Hubbard model is
introduced and it is shown how the temperature dependence of bond orders
is related to the thermal expansion. We also discuss how the
parameters in the Hubbard model depend on the lattice constants.
Results of calculations of the bond orders using the Finite
Temperature Lanczos Method are presented in Section \ref{sec:results}.
Comparisons are made between the calculated thermal expansion (for a
range of parameter values) and specific experiments on organic charge
transfer salts.  This is followed by discussion of remaining future
challenges, while we summarize our main conclusions in Section
\ref{sec_conc}.

\section{General thermodynamic considerations}
\label{sec:thermo}

For simplicity and to elucidate the essential physics we first discuss
the isotropic case.  Experiments are done at constant temperature,
pressure, and particle number (assuming the sample is not connected to
electrical leads and the particle density is controlled by chemistry).
Thus, the Gibbs free energy $G(T,P,N_e)$ is a minimum and satisfies
 \begin{equation}
dG=-SdT + VdP + \mu dN_e.
\end{equation}
From this we can derive a Maxwell relation implying that the volume
thermal expansion is given by
 \begin{equation}
\alpha(T) \equiv {1 \over V} \left({ \partial V \over \partial T}\right)_P
= -{1 \over V} \left({ \partial S \over \partial P}\right)_T.
\end{equation}
In calculations it is however easier to vary the volume than pressure
and so we rewrite this as 
 \begin{equation}
\alpha(T) 
= -{1 \over V} \left({ \partial V \over \partial P}\right)_T
\left({ \partial S \over \partial V}\right)_T
=\kappa_T \left({ \partial S \over \partial V}\right)_T ,
\label{alphat}
\end{equation}
where $\kappa_T$ is the isothermal bulk compressibility.  Given the
expression above it is natural to expect strong thermal expansion
effects when the entropy is large (e.g., at the incoherence crossover)
and sensitive to volume-dependent parameters in the system (e.g.,
close to the Mott transition).

As the volume changes so do the lattice constants and the
parameters in an underlying electronic Hamiltonian such as the hopping
integral $t$ in a Hubbard model.  To evaluate (\ref{alphat}) we use
\begin{equation}
\left({ \partial S \over \partial V}\right)_{T,N_e}
=
{ \partial t \over \partial V}
\left({ \partial S \over \partial t}\right)_{T,N_e} ,
\label{alphat2}
\end{equation}
leading to 
 \begin{equation}
\alpha(T) 
=\kappa_T 
{ \partial t \over \partial V}
\left({ \partial S \over \partial t}\right)_{T,N_e}.
\label{alphat3}
\end{equation}
This equation for volume thermal expansion applies for isotropic case,
while orientational dependance of thermal expansion can be obtained in
a similar manner by generalizing  $VdP$ to 
$-\sum_i d\sigma_i V^0 l_i/l_i^0$. Here $i=x,y,z$, $d\sigma_i$ is
the change of uniaxial stress, $l_i$ is the length in $i$ direction, while
$V^0$ and $l_i^0$ are reference volume and length. 
Thermal expansion in direction $i$ is then, similarly as
Eq. \ref{alphat2}, given by
 \begin{equation}
\alpha_i(T) 
=\frac{1}{E_i}
\frac{l_i^0}{V^0}
{ \partial t \over \partial l_i}
\left({ \partial S \over \partial t}\right)_{T,N_e} 
+
\cdots
\label{alphait}
\end{equation}
Additional terms involve different electronic model parameters instead
of $t$, e.g., $t'$ and $U$. This expression is valid for small
Poisson's ratio, which is together with more detailed derivation in
terms of the grand potential ($\Omega=G-PV-\mu N_e$) discussed in
Appendix \ref{sec_app_anis}.  The value of the Young's modulus $E_i$ we
take from experiment and later comment on the effect of the Mott
transition on it. We estimate $\partial t/\partial l_i$  from band
structure calculations and we calculate $\partial S/\partial t$ 
numerically with the Finite Temperature Lanczos Method (FTLM)
\cite{jaklic00,kokalj13}.  It follows from the third law of
thermodynamics that $\alpha_i(T) \to 0$ as $T \to 0$.


\section{Hubbard model}
\label{sec:hubbard}


We model our system with two Hamiltonian terms, $\hat H=\hat
H_{el}+\hat H_{other}$, 
where $\hat H_{el}$ describes electrons in the highest occupied band
and their contribution to the thermal expansion is our main
interest. We decouple these electronic degrees of freedom from
others such as phonons and electrons in lower filled bands, and denote
their contribution with $\hat H_{other}$. With this we neglected the
direct coupling of electrons and phonons, but we keep 
the dependence of $\hat H_{el}$ on
the lattice constants $a_i$, which is in the spirit of a Born-Oppenheimer
approximation.



We model the electrons in the highest occupied band
with the grand canonical Hubbard model on the anisotropic triangular
lattice,  
\eqa{
  \hat H_{el}&=&-\sum_{i,j,\sigma} t_{i,j} c_{i,\sigma}^\dagger c_{j,\sigma} +U\sum_i
  \hat n_{i,\uparrow} \hat n_{i,\downarrow} 
-\mu\sum_{i,\sigma} \hat n_{i,\sigma}\nonumber\\
&=&
-t \hat T_1 -t' \hat T_2+U \hat D-
\mu \hat N_e .
\label{eq_hel}}
Here $t_{i,j}=t$ for nearest neighbor bonds in two directions and
$t_{i,j}=t'$ for nearest neighbor bonds in the third direction
(compare Figure \ref{fig_0}). Electronic spin is denoted with $\sigma$
($\uparrow$ or $\downarrow$).  $\hat T_1$ and $\hat T_2$ denote bond
order operators corresponding to $t$ and $t'$ hopping respectively,
and $\hat D$ is the double occupancy operator.
The chemical potential $\mu(T)$ is determined by the
required half filling, i.e. that $\langle \hat N_e \rangle = N_e = N$,
where $N$ is the 
number of lattice sites. $\aver{\ldots}$ denotes thermal average.

Following equation (\ref{alphait}), we relate the thermal
expansion to   
$\left({ \partial S \over \partial x}\right)_{T,N_e}$ where $x=t,t',U$.
Using the Maxwell-type relations we can write
\eqa{
\left({ \partial S \over \partial t}\right)_{T,N_e,t',U} 
=\left({ \partial  \langle \hat T_1\rangle \over \partial T}\right)_{N_e,t,t',U},
\label{dsdt}\\
\left({ \partial S \over \partial t'}\right)_{T,N_e,t,U} 
=\left({ \partial  \langle \hat T_2\rangle \over \partial T}\right)_{N_e,t,t',U},
\label{dsdt’}\\
\left({ \partial S \over \partial U}\right)_{T,N_e,t,t'}  
=-\left({ \partial  \langle \hat D\rangle \over \partial T}\right)_{N_e,t,t',U}.
\label{dsdu}
}
With the equations above we related the thermal expansion to the variation of
entropy with electronic model parameters ($t, t'$ and $U$) or
analogously to the $T$ dependence of bond orders ($\aver{\hat T_1}$,
$\aver{\hat T_2}$) or double occupancy ($\aver{\hat D}$), again at
fixed particle number.

\subsection{Dependence of the Hubbard model parameters on the lattice
  constants }   

The expression (\ref{alphait}) 
for the thermal expansion requires knowledge of the
dependence of Hubbard model parameters ($t$, $t'$ and $U$) on the lattice
constants. Estimates of this dependence can be obtained from
electronic structure calculations via methods such as
extended H\"uckel or Density Functional Theory (DFT).
Calculations using the former with the 
experimental crystal structure for 
X=Cu$_2$(CN)$_3$ 
give the following (compare
Fig. 8 in Ref. \onlinecite{shimizu11}), 
\eqa{
  t=t_0(1-4.9(\frac{c-c_0}{c_0})), \label{eq_teqt0} \\
  t'=t'_0(1-8.7 (\frac{b-b_0}{b_0})). \label{eq_tpeqtp0}\\
U = U_0(1- 3.5(\frac{b-b_0}{b_0})- 2.8(\frac{c-c_0}{c_0})). \label{eq_UeqU0}
}

Here $c$ and $b$ are in-plane lattice constants (compare Figure
\ref{fig_0}), while reference values at 1 bar pressure are denoted
with $c_0, b_0, t_0$, $t'_0$ and $U_0$. In general the Hubbard model
parameters depend on all lattice constants and structural
parameters (including angles)
\cite{mori98,mori99,kondo03,kandpal09,jeschke12} and all should be
considered, but for simplicity we keep only the dependencies given
above.  They were obtained\cite{shimizu11} by assuming that squeezing only reduces the
intermolecular distance along the direction of the uniaxial stress,
but does not induce rotations of molecules.

The actual dependence of the Hubbard $U$ on the structure is subtle.
In a crystal such as sodium or nickel oxide $U$ would simply be
associated with a single atomic orbital and would not vary with
lattice constant and stress, provided screening is
neglected. Screening could introduce some dependence.  However, for
(BEDT-TTF)$_2$X crystals things are more complicated because $U$ is
with respect to a molecular orbital on a dimer of BEDT-TTF molecules
and the dimer geometry will vary with uniaxial stress.  Furthermore,
the estimate given in the expression (\ref{eq_UeqU0}) is based on the
assumption that $U$ is solely given by the intradimer hopping integral
$2t_{b1}$.  However, that is only in the limit $t_{b1} \ll \tilde U_0$, where
$\tilde U_0$ is the Coulomb repulsion associated with single BEDT-TTF
molecule. Although, this assumption is actually unrealistic
\cite{powell11}, Eq. (\ref{eq_UeqU0}) is useful as an estimate,
particularly because it is an upper bound for the dependence.
Furthermore, we will see below that the $U$ dependence of the entropy
is much smaller than that of the $t$ and $t'$ dependence (compare
Figs. \ref{fig_2} and \ref{fig_3}) and so turns out to make a
relatively insignificant contribution to the thermal expansion. Hence,
the above concerns are not particularly important.

The thermal expansion in the direction of the $c$ axis can therefore
be calculated from Eqs. \eqref{alphait}, \eqref{dsdt}, and \eqref{eq_teqt0}, 
 related to the bond order $\aver{\hat T_1}$, to give
\eq{
\alpha_c=\frac{1}{E_c}\frac{c_0}{N V_{1uc}} 
\frac{\partial t}{\partial c}
  \frac{\partial\aver{\hat T_1}}{\partial T}
\label{eq_alphac}
}
where $E_c$ is the Young's modulus in the $c$ direction, and 
we have neglected the contribution from the double occupancy.
$V_{1uc}$ denotes the volume of one unit cell. 
In a similar way, the main contribution to $\alpha_b$ is given by the $t'$
dependence on $b$ according to Eq. \eqref{eq_tpeqtp0} and  
the temperature derivative of $\aver{\hat T_2}$, 
\eq{
\alpha_b=\frac{1}{E_b}\frac{b_0}{N V_{1uc}} 
\frac{\partial t'}{\partial b}
  \frac{\partial\aver{\hat T_2}}{\partial T}.
\label{eq_alphab}
}
These are the expressions we use below to calculate the thermal
expansion. 

\section{Results}
\label{sec:results}

In the following we discuss several numerical results obtained on
$N=16$ sites by the finite-temperature Lanczos method (FLTM)
\cite{jaklic00}, which was previously successfully used to determine a
range of thermodynamic quantities for the same Hubbard model
\cite{kokalj13}.  In particular, it was shown that one could describe
the Mott metal-insulator transition and the crossover from a Fermi
liquid to a bad metal.

\subsection{Dependence of the entropy on Hubbard model parameters}

\begin{figure}[htb] 
 \centering 
\includegraphics[width =52mm,angle=-90]
  {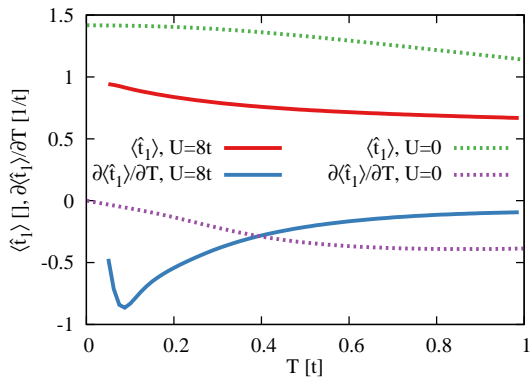}
  \caption{
(color online) Strong correlations significantly enhance
    the electronic contribution to the thermal expansion through the
    temperature dependence of the bond order. This is demonstrated by
    comparing $\partial \aver{\hat t_1}/\partial T$ at low $T$ for
    $U=8t$ with the noninteracting $U=0$ result. $\aver{\hat
      t_1}=\aver{\hat T_1}/N$ is the average kinetic energy in certain
    directions.  At low temperatures the enhancement is by an order of
    magnitude.  The plotted quantity is related to the thermal
    expansion via Eq. \eqref{eq_alphac} and to the stress dependence of
    entropy $s=S/N$ via a Maxwell relation $\partial \aver{\hat
      t_1}/\partial T=\partial s/\partial t$, Eq. \eqref{dsdt}.  In addition, strong
    correlations also produce a strong non-monotonic temperature
    dependence. Results are for $t'=0.8t$ and $U=8t$, corresponding
    to the system in the Mott insulating phase.\cite{kokalj13} }
\label{fig_1}
\end{figure}

In Fig. \ref{fig_1} we show the $T$-derivative of $\aver{\hat
  t_1}=\aver{\hat T_1}/N$, namely $(1/N)(\partial \aver{\hat
  T_1}/\partial T)_{t,t',U,N_e}$, in the insulating phase ($U=8t$,
$t'=0.8t$ \cite{kokalj13}) and compare it to the result for
noninteracting fermions ($U=0$). The strong difference shows that
correlations can increase the electronic contribution to the thermal
expansion by as much as an order of magnitude at low temperatures, and
produce a non-monotonic temperature dependence.


\begin{figure}[htb] 
 \centering 
\includegraphics[width =52mm,angle=-90]
  {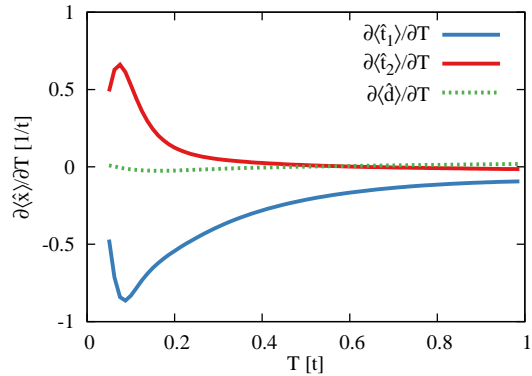}
  \caption{(color online) Strongly anisotropic temperature dependence of the bond
    orders $\aver{\hat T_1}$ and $\aver{\hat T_2}$. Due to strong correlations and frustration
    a small anisotropy (i.e. deviation from the isotropy of
the triangular lattice) with
    $t'=0.8t$ leads to strongly anisotropic electronic contributions to
    the thermal expansion. This is seen by comparing 
$\partial \aver{\hat t_1}/\partial T=\partial s/\partial t$ and 
$\partial \aver{\hat t_2}/\partial T = \partial s/\partial t'$, which
    are large and have opposite sign at low $T$. For the isotropic case
    ($t'=t$) they have essentially the same $T$ dependence, with only a
    factor of two difference coming from a number of bonds associated with
    corresponding hopping. The double occupancy $\aver{\hat d}=\aver{\hat
      D}/N $ 
shows a much weaker $T$ dependence.
 Results    are for the insulating phase with $t'=0.8t$ and $U=8t$ \cite{kokalj13}. 
}
\label{fig_2}
\end{figure}

In Fig. \ref{fig_2} we show that an anisotropy value of
$t'/t=~0.8$ leads to strong anisotropy of bond orders and their
$T$-derivative relevant for thermal expansion. This probably
originates in strong frustration for the isotropic case with large low
$T$ entropy and therefore small changes
in the anisotropy can lead to strong change of
bond orders which in the insulating phase 
are associated with spin correlations.  In
Fig. \ref{fig_2} we also show the $T$-derivative of double occupancy which
has smaller values than for bond orders.
By the Maxwell relation in Eq.~\eqref{dsdu},
our results in
Fig. \ref{fig_2} are qualitatively
consistent with the $U$ variation of $S$ shown in
Fig. 4 of Ref.  \onlinecite{kokalj13}. 
This relation of the
entropy and negative values of $(\partial \aver{\hat D}/\partial
T)_{t,t',N_e}$ at low $T$ were recently evoked \cite{li14,laubach14} as a
possible mechanism for adiabatic cooling in optical lattices.


\begin{figure}[htb] 
 \centering 
\includegraphics[width =52mm,angle=-90]
  {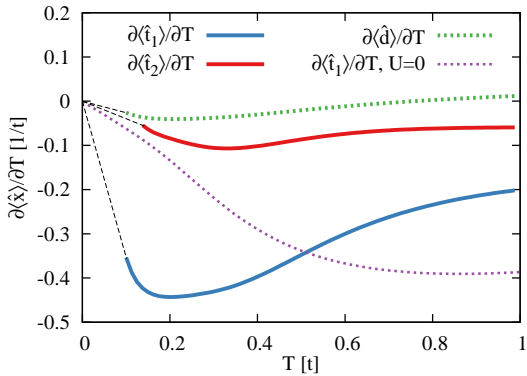}
  \caption{(color online) In the metallic phase ($t'=0.8t$, $U=6t$
    \cite{kokalj13}) the temperature dependence of the bond orders is
    strongly anisotropic and non-monotonic.  At low temperatures, a
    linear dependence is expected for a Fermi liquid, extrapolating to
    zero, in accordance with the third law of thermodynamics.  This is
    shown by the dashed lines.  The linearity ceases above the
    coherence temperature $T_\textrm{coh}$, where there is a crossover to a
    bad metal and where a maximum magnitude of the thermal expansion
    is observed.  At low temperatures ($T \simeq 0.1t$) $\partial
    \aver{\hat t_1}/\partial T= \partial s/\partial t$ can be an order
    of magnitude larger than for the non-interacting ($U=0$) system.
  }
\label{fig_3}
\end{figure}

In Fig. \ref{fig_3} we show results for a metallic case ($t'=0.8t$,
$U=6t$ \cite{kokalj13}) for which a Fermi liquid like behaviour is
expected at low $T$ leading to
a linear-in-$T$ thermal expansion coefficient
below the coherence temperature $T_\textrm{coh}$, above which a crossover to a
bad metallic phase appears \cite{merino00}. 
Such a linear in $T$ dependence originates in
$\alpha \propto - \partial S/\partial x$ (with $x=t$, $t'$ or $U$)
and a linear-in-$T$ entropy $S$ . Such dependence of entropy and its
variation with $U$ is shown in Fig. 4 in Ref. \onlinecite{kokalj13}.
Based on these considerations we include in Figs. \ref{fig_3} and
\ref{fig_5} a linear extrapolation of the FTLM results to zero
temperature.


\subsection{Thermal expansion coefficients}

We now present the results of calculations that
can be compared to experimental data for the  thermal expansion of 
specific organic charge transfer salts.
We used Eqs. (\ref{eq_alphac}, \ref{eq_alphab})
 together with the
following estimates for parameter values: $V_{1uc}=800\times 10^{-30}
$ m$^3$ from Fig. 5 in Ref. \onlinecite{jeschke12}, the temperature
scale is determined by $t=50$~meV \cite{kokalj13}, estimated
from Density Functional Theory (DFT)-based
calculations \cite{kandpal09, nakamura09, jeschke12}.  Estimates for
the Young's modulus from X-ray determination of the crystal structure
under uniaxial stress are $1/E_c=(1/c_0)(dc/d\sigma_c)=6.9\times
10^{-11}$ Pa$^{-1}$ and $1/E_b=(1/b_0)(db/d\sigma_b)=5\times 10^{-11}$
Pa$^{-1}$ from Table 1 in Ref. \onlinecite{kondo03} for
$\alpha$-(BEDT-TTF)$_2$NH$_4$Hg(SCN)$_4$.  Comparable values for
isotropic pressure in $\kappa$-(BEDT-TTF)$_2$Cu(NCS)$_2$ are given in
Ref. \onlinecite{rahal1997}. We also use Eqs. (\ref{eq_teqt0},
\ref{eq_tpeqtp0}) for estimates of $\partial t/\partial c$ and
$\partial t'/\partial b$.

\begin{figure}[htb] 
 \centering 
\includegraphics[width =52mm,angle=-90]
{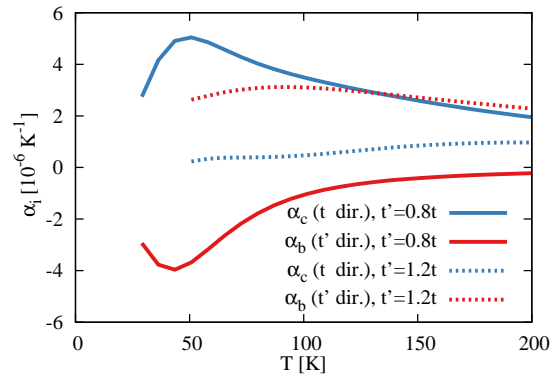}
\caption{(color online) Temperature dependence of the thermal
  expansion in the Mott insulating phase.  Note the non-monotonic
  behavior and the large anisotropy. Indeed, in the $b$ direction,
  thermal contraction rather than expansion occurs.  The maximum
  magnitude occurs at approximately the same temperature as that for
  which the specific heat and magnetic susceptibility are a maximum
  (compare Figure S1 of Ref. \onlinecite{kokalj13}).  The solid curves
  ($t'=0.8t$ and $U=8t$) can be compared to experimental results for
  $\kappa$-(BEDT-TTF)$_2$Cu$_2$(CN)$_3$ shown in Fig. 1 in
  Ref. \onlinecite{manna10}.  The results are quite sensitive to the
  parameter values and proximity to the Mott transition.  This is seen
  by comparing the dashed curves ($t'=1.2t$).  The parameter values
  used are described in the text.  }
\label{fig_4}
\end{figure}

In Fig. \ref{fig_4} we show an estimate of the thermal expansion
coefficients for the insulating phase and parameters ($t'=0.8t$,
$U=8t$) that correspond to $\kappa$-(BEDT-TTF)$_2$Cu$_2$(CN)$_3$, and
can be compared to experimental data shown in Fig. 1 of
Ref. \onlinecite{manna10}.  The calculated magnitude of about $5
\times 10^{-6}$/K at 50 K is approximately one fifth of the 
measured value.  We discuss possible
explanations of this discrepancy later.
As in experiment we observe a strong anisotropy with maximum around 50
K, but the sign of the anisotropy is opposite to the experimental one at
such $T$. Interestingly, a similar $T$ dependence with the right
absolute values is experimentally observed as a very low $T$ ($\sim 6$
K) anomaly (see Fig. 2 in Ref. \onlinecite{manna10}), but for
agreement our $T$ scale would need to be reduced by a factor of 10,
suggesting that this involves different physics beyond our
calculations, such as transition into some type of spin liquid phase.

Our results in Fig. \ref{fig_4} have significantly different
$T$-dependencies for the  thermal expansion coefficients in $c$ ($t$) and
$b$ ($t'$) directions due to anisotropy in
the bond orders  shown in
Fig. \ref{fig_2}, originating in the anisotropy $t'=0.8t$ and since variation 
of the different lattice constant changes differently $t$ and $t'$
(Eqs. \ref{eq_teqt0}, \ref{eq_tpeqtp0}).  Low temperature
 experimental results
shown in Fig. 1 of Ref. \onlinecite{manna10} show a strong difference in
the $T$-dependence between the $b$ and $c$ directions, suggesting that, if
they originate from the electronic degrees of freedom, the proper
electronic model should have notable $t$-$t'$ asymmetry, or that
the dependence of $t$ and $t'$ on the lattice constants $c$ and $b$ is
strongly asymmetric. The anisotropy $\alpha_c-\alpha_b$ in our results
shown in Fig. \ref{fig_4} has the opposite sign to experiment.  Taking
$t' \sim 1.2t > t$ changes the sign of our $\alpha_c-\alpha_b$
results, making the comparison to experiment better. Change of the
sign of the thermal expansion
by increasing $t'$ above the isotropic point ($t'=t$) originates
in moving away from maximal frustration (and therefore maximal
entropy).  This also involves moving away from the isotropic point for
which the temperature dependence of both $\aver{\hat T_1}$ and
$\aver{\hat T_2}$ is essentially the same (apart from a factor of 2)
due to symmetry (compare Figure \ref{fig_2}).

\begin{figure}[htb] 
 \centering 
\includegraphics[width =52mm,angle=-90]
  {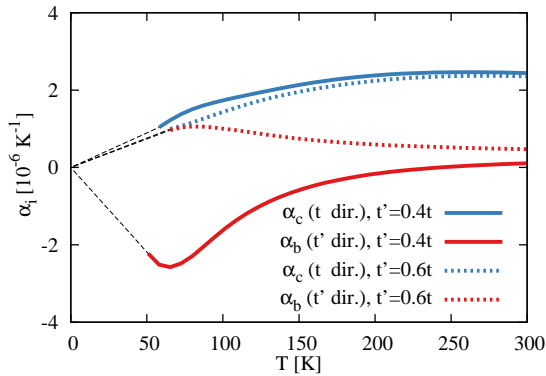}
  \caption{(color online) Temperature dependence of the thermal
    expansion in the metallic phase ($U=4t$).  These results can be
    compared to Fig. 5a and 5b in Ref. \onlinecite{muller02} with the
    $t'=0.4t$ results more relevant for Fig. 5a ($\kappa$-Br) and
    $t'=0.6t$ results more relevant for Fig. 5b ($\kappa$-NCS).  The
    dashed lines are linear extrapolations to zero temperature, as
    expected for a Fermi liquid.  Note that $t'=0.4t$ is closer to the
    Mott insulating phase than $t'=0.6t$ (compare Fig. 3 in
    Ref. \onlinecite{kokalj13}).  Our $\alpha_c$ ($\alpha_b$) should
    be compared to the experimental data shown as full squares (empty
    circles) in Fig. 5 in Ref. \onlinecite{muller02}. Our calculated
    values are about 5-10 times smaller than the measured values. As
    in experiment we observe for $T\sim60$ K larger anisotropy
    ($\alpha_c-\alpha_b$) for $t'=0.4t$ than for $t'=0.6t$ with the
    right sign for $t'=0.4t$.  We observe a maximum magnitude at
    $T\sim T_\textrm{coh}\sim 60$ K, suggesting that the experimental
    anomalies at $T\sim 50$ K could have an electronic origin,
    although the observed increase (decrease) of $\alpha_c$ for
    $\kappa$-Br ($\kappa$-NCS) at such temperature is inconsistent
    with our results.  We use the same parameter values and
    approximations as for Fig. \ref{fig_4}.  }
\label{fig_5}
\end{figure}

In Fig. \ref{fig_5} we show our estimate of the electronic contribution to
the thermal expansion for the metallic phase of organic charge
transfer salts.
Similar to the experimental data, our results show a maximum at $T\sim
60$ K and suggest that the experimentally observed anomalies (see
Fig. 5 in Ref. \onlinecite{muller02}) could have an electronic origin.
On the other hand, in Fig. \ref{fig_5} we observe larger anisotropy
$(\alpha_c-\alpha_b)$ for $t'=0.4t$ than for $t'=0.6t$, which is in
agreement with experimentally observed larger ($\alpha_a-\alpha_c$)
for $\kappa$-(H$_8$-ET)$_2$Cu[N(CN)$_2$]Br ($\kappa$-Br) shown in
Fig. 5a in Ref. \onlinecite{muller02} than ($\alpha_c-\alpha_b$) for
$\kappa$-(D$_8$-ET)$_2$Cu(NCS)$_2$ ($\kappa$-NCS) shown in Fig. 5b in
Ref. \onlinecite{muller02}. We note that in
$\kappa$-(H$_8$-ET)$_2$Cu[N(CN)$_2$]Br $t'\sim0.4t$ while in
$\kappa$-(D$_8$-ET)$_2$Cu(NCS)$_2$ $t'\sim0.6t$ \cite{kandpal09}.

\subsection{ Sign of the hopping integrals}

We note that for comparison with the organics in
Figs. \ref{fig_4}  and \ref{fig_5} we used positive $t$ and $t'$ for the Hubbard
model defined by Eq. \eqref{eq_hel}, while with
respect to our definition DFT based calculations
suggest they are both negative \cite{kandpal09}. This is not a problem,
since changing the signs of both $t$ and $t'$ corresponds at half-filling to a
particle-hole transformation and leads to the same result due to a double
sign change, 
e.g., $\partial \aver{\hat T_1}/\partial T \to - \partial \aver{\hat
  T_1}/\partial T$ and $\partial t/\partial c \to -\partial t/\partial
c$. On the other hand, Refs. \onlinecite{nakamura09,koretsune14} suggest
negative $t'/t$ which could affect the results but actually also
$t'/t\to-t'/t$ corresponds to a particle-hole transformation with an
additional $k$-space shift of $(\pi,\pi)$ (considering the
equivalent square lattice with one 
diagonal hopping $t’$). This again does not change the results for
thermal expansion. On the other hand, such particle-hole
transformations are important for the sign of the thermopower
\cite{kokalj14}. 


\subsection{Future challenges}

We now discuss several possible improvements to our theoretical
description that we leave as 
future challenges. First it is clear from discussion of
Eq. \eqref{alphait} and furthermore from discussion of anisotropic
effects in Appendix \ref{sec_app_anis} and Eq. \eqref{eq_alphaieqsumij}
therein that for an anisotropic materials like the organics the
stiffness tensor $C_{ijkl}$ can be strongly anisotropic with several
important elastic constants that are not known at a moment, but may be
experimentally accessible. For example, adding Poisson's ratios
to known Young's moduli and extracting also other elastic constants
would allow for a full tensor description. This is not just of interest
for the study of thermal expansion, but also on its own, since also stiffness
tensor has notable electronic contributions.  These have been already
observed as lattice softening, e.g., via sound velocity
\cite{fournier03, hassan05},
which becomes substantial (up to 50\%)
close to the metal-insulator transition (MIT) and in addition suggests
critical behaviour at the end of the first-order line, leading to a
diverging $\partial^2 \Omega/\partial t^2$ related to
$\partial^2\Omega/\partial l_i^2$ [see Eq. \eqref{eq_cij}]. One should
however keep in mind, that the MIT is experimentally observed to be
weakly first order in the organics \cite{limelette03} but its order in
a Hubbard model at half filling is still controversial
\cite{yamada14,yoshioka09,mishmash14,yang10}.
In our analysis we do not include these
lattice softening effects (reduced Young's modulus) close to the
metal-insulator transition, but their inclusion would increase our
$\alpha_i$ by roughly a factor of two, making the electronic contribution
to $\alpha_i$ larger and more important and would improve the
comparison to experiment (see in particular the discussion of
Fig. \ref{fig_5}).

Another challenge is to obtain the dependence of the Hubbard model
parameters ($t$, $t'$ and $U$) on all lattice constants $a_i$ and on
all structural parameters, including the angles and orientation of
molecules.  These dependencies are not easy to obtain, and simple
Eqs. (\ref{eq_teqt0}-\ref{eq_UeqU0}) could be greatly
improved with more elaborate DFT calculations or studies such as in
Refs. \onlinecite{mori98,mori99}. In particular Fig. 2 in
Ref. \onlinecite{mori99} shows that in various salts the different
angle between ET molecules is directly connected to the lattice
constants, which suggests that this angle is an important structural
parameter and that it possibly varies also with temperature and
applied stress. Therefore DFT calculations, which would in addition to
intermolecular spacing relax also angles between molecules, could be
valuable and present a future challenge.  DFT could connect changes of
$H_{el}$ parameters to changes of structural parameters with the
complete tensor.  This would further facilitate the full tensor
description of electronic contribution to thermal expansion and
elastic constants.


\section{Conclusions}
\label{sec_conc}

We have shown how the electronic contribution to the thermal
expansion is related to the electronic degrees of freedom via the
parameters ($t$, $t'$ and $U$) in
a Hubbard model and temperature derivatives of known
quantities (bond orders and double occupancy). The values of thermal
expansion coefficients are further governed by the relation of model
parameters to lattice (structural) constants and by elasticity
constants. 

The electronic contribution to the thermal expansion is large with
strong orientational and non-monotonic temperature
dependence. Furthermore, we showed that correlations strongly increase
the electronic contribution and by estimating it for organic charge
transfer salts we showed that it can provide a qualitative
understanding of experimental data for temperatures below 100 K.  In
particular, contrary to suggestions in Ref. \onlinecite{muller02}, the anomalies
around 50 K may not be lattice anomalies or structural phase
transitions, rather they could originate from the electronic
contribution, and be due to the bad metal - Fermi liquid crossover.

It should be stressed that also phononic contribution to the thermal
expansion may play an important role at quite low $T$, which is
suggested by large phononic contribution to specific heat (see
Ref. \onlinecite{yamashita08,yamashita11} and the Supplement of
Ref. \onlinecite{kokalj13}) and in turn to entropy relevant 
for thermal expansion [Eq. \eqref{alphat}]. 
Therefore the study of lattice vibrations (e.g., anharmonic
effects, orientational dependence or the Gr\"uneisen parameter
\cite{ashcroft}) and the estimates of their contribution to thermal
expansion or stiffness tensor may aid our
understanding.

\begin{acknowledgments}
  We acknowledge helpful discussions with H.O. Jeschke, P. Prelov\v{s}ek, and R. Valent\'i. 
This work was supported by Slovenian Research Agency grant
  Z1-5442 (J.K.) and an Australian Research Council Discovery Project
  grant (R.H.M.). 
\end{acknowledgments}


\appendix
\section{Anisotropic thermal expansion}
\label{sec_app_anis}

We discuss thermal expansion here in terms of a grand
potential $\Omega$, due to its simple connection to the 
electronic Hamiltonian
\eq{
\exp{(-\beta \Omega)}=\tr [\exp{(-\beta \hat H_{el})}],
}
and its straight forward calculation within FTLM \cite{jaklic00}.
Thermal expansion coefficients are given by
\eq{
\alpha_i \equiv \frac{1}{l_i} \left (\frac{\partial l_i}{\partial
    T}\right)_{P,N_e}, 
}
where $l_i$ is a length of a sample in the $i$ ($=x, y$ or $z$) direction
and can be exchanged also by a lattice constant $a_i$, and where we
denoted that experiments are done at constant pressure ($P$) and fixed
electron number ($N_e$). Since we are interested also in an
orientational ($i$) dependence, we first need to generalize the standard
mechanical work $-PdV$ to $V^0\sum_{i,j} \sigma_{ij}d\varepsilon_{ij}$
with $V^0$ being a reference volume, while $\sigma_{ij}$ and 
$\varepsilon_{ij}$ are stress and strain 
tensors, respectively. 
We however simplify our analysis by considering just normal
stress and no shear deformations taking only diagonal terms.
$\sigma_{ii}=\sigma_i$ is uniaxial stress which equal $-P$ for
isotropic pressure and $\varepsilon_{ii}=dl_i/l_i^0$ with $l_i^0$
denoting reference length.
With this we can write
mechanical work as $\sum_i \sigma_i V^0 dl_i/l_i^0$. This brings us to
$\Omega=\Omega(T,l_i,\mu)$ and $d\Omega= -SdT+\sum_i 
\sigma_i V^0 dl_i/l_i^0 -N_e d\mu$, where for a fixed $N_e$ one needs to
adjust the chemical potential, $\mu=\mu(T,l_i,N_e)$. 
From $\Omega$ one can obtain the equation of state which for usual
work ($-PdV$)  reads $-P=(\partial \Omega/\partial V)_{T,\mu}$ but with
our generalized work the three equations of state (for $i=x,y,z$) are
\eq{
\sigma_i=\frac{l_i^0}{V^0}\left (\frac{\partial \Omega}{\partial l_i} \right
)_{T,l_{j\ne i}, \mu}.
\label{eq_sigmaisi}
}
Taking the full derivative of equation of state for fixed $N_e$ in the
case of usual work ($-PdV$) one obtains differential equation of state
$
-dP=
(\frac{\partial }{\partial T} 
(
\frac{\partial \Omega}{\partial V}
 )_{T,\mu}
 )_{V,N_e} 
dT
+
(\frac{\partial }{\partial V} 
(
\frac{\partial \Omega}{\partial V}
 )_{T,\mu}
 )_{T,N_e}
dV
$, 
which when compared to $dV/V=\beta dT-\kappa_T dP$ gives expression for
isothermal bulk compressibility 
$\kappa_T^{-1}=
V
 (\frac{\partial }{\partial V} 
(
\frac{\partial \Omega}{\partial V}
 )_{T,\mu}
 )_{T,N_e}
$
and volume thermal expansion
 $\beta = \kappa_T
( \frac{\partial }{\partial T} 
(
\frac{\partial \Omega}{\partial V}
 )_{T,\mu}
 )_{V,N_e} $
in terms of $\Omega$. Similarly taking full differentials of
Eq. \eqref{eq_sigmaisi} leads to differential equations of states
\eqa{
d\sigma_i&\!\!=&\!\!\frac{l_i^0}{V^0} 
\Big (
\frac{\partial}{\partial T}\!
\Big(
\frac{\partial \Omega}{\partial l_i}
\Big )_{\!T,l_{k\ne i},\mu}
\Big )_{\!l_j,N_e}
\!dT
\!+\!\sum_jC_{ij}\frac{dl_j}{l_j^0},\label{eq_dsigmai}\\
C_{ij}&\!\!=&\!\!
\frac{l_i^0 l_j^0}{V^0} 
\Big (
\frac{\partial}{\partial l_j}\!
\Big(
\frac{\partial \Omega}{\partial l_i}
\Big )_{\!T,l_{k\ne i},\mu}
\Big )_{\!T,l_{k\ne j},N_e}.\label{eq_cij}
} 
From above Eq. \eqref{eq_dsigmai} it is clear that 
a small change of strain $dl_j/l_j^0=\varepsilon_j$ leads to a small
change of stress $d\sigma_i=\tilde \sigma_i$, which are at  constant
temperature ($dT=0$) related by $\tilde \sigma_i=C_{ij}\varepsilon_j$ or
with expanded indices $\tilde \sigma_{ii}=C_{iijj}\varepsilon_{jj}$,
namely by Hook's law. 
Now we recognise $C_{ij}$ or $C_{iijj}$ as a stiffness tensor, which
depends on material's elastic constants,
and has replaced $\kappa_T^{-1}$.  The symmetry of $C_{ij}$ is
discussed in Appendix \ref{sec_symmetry_cij}.

Thermal expansion coefficients can now be expressed as
\eq{
\alpha_i=\sum_{j}
(C^{-1})_{i,j} 
\frac{-l_j^0}{V^0 } 
\Big (
\frac{\partial}{\partial T}
\Big(
\frac{\partial \Omega}{\partial l_j}
\Big )_{T,l_{k\ne j},\mu}
\Big )_{l_k,N_e},
\label{eq_alphaieqsumij}
}
and we further for clarity simplify our calculation by assuming that
Poisson's ratio is small which makes 
$C^{-1}$ diagonal, $(C^{-1})_{i,j}=(1/E_i) \delta_{ij}$ with $E_i$
being Young's modulus in $i$ direction. 

Similarly one can
show that $l_i$- and $T$-derivatives of $\Omega$
in Eq. \eqref{eq_alphaieqsumij}, can be replaced with $l_i$ derivative
of entropy $S$. See Appendix \ref{sec_texpan_entropy} for more detail. 
\eq{
\alpha_i=\frac{1}{E_i}
\frac{l_i^0}{V^0} 
\Big(
\frac{\partial S}{\partial l_i}
\Big )_{T,l_{k\ne i},N_e}.
\label{eq_alphaieqei}
}

 Further more, since $E_i>0$ the sign
of $\alpha_i$ is determined by the entropy derivative and therefore
whether the change of $l_i$ (or in turn some electronic model parameter,
see Eqs. \eqref{eq_tpeqtp0} and \eqref{eq_teqt0}) increases or decreases the entropy. For maximally
frustrated systems the low-$T$ entropy is expected to be maximal and therefore the
sign of $\alpha_i$ can help determining  whether one is with a certain 
parameter above or below the maximal frustration.

\section{Relation of thermal expansion to entropy via grand
  potential} 
\label{sec_texpan_entropy}
Here we show that the $T$ and $l_i$ derivative of $\Omega$, one at
fixed $N_e$ and the other at fixed $\mu$, appearing in
Eq. \eqref{eq_alphaieqsumij} for thermal expansion can be expressed as
$l_i$ derivative of entropy. Such relation can be shown with the
use of Helmholtz free energy $F$, but here we show it by using $\Omega$. 
\eqa{
\Big (
\frac{\partial}{\partial T}
\Big(
\frac{\partial \Omega}{\partial l_i}
\Big )_{T,l_{k\ne i},\mu}
\Big )_{l_k,N_e}
&\!\!=&\!\! \Big(
\frac{\partial ^2 \Omega}{\partial T\partial l_i}
\Big)_{l_{k\ne i},\mu}\nonumber\\
&\!\!+&\!\!
\Big(
\frac{\partial ^2 \Omega}{\partial \mu\partial l_i}
\Big)_{T, l_{k\ne i}}
\Big(
\frac{\partial  \mu}{\partial T}
\Big)_{ l_{k},N_e}\label{eq_ddtdomegadli}\\
-\Big(
\frac{\partial S}{\partial l_i}
\Big )_{T,l_{k\ne i},N_e}
&\!\!=&\!\! \Big(
\frac{\partial ^2 \Omega}{\partial l_i\partial T}
\Big)_{l_{k\ne i},\mu}\nonumber\\
&\!\!+&\!\!
\Big(
\frac{\partial ^2 \Omega}{\partial \mu\partial T}
\Big)_{l_{k}}
\Big(
\frac{\partial  \mu}{\partial l_i}
\Big)_{T, l_{k\ne i},N_e}.\label{eq_dsdli}
}
Since $-N_e=(\partial \Omega /\partial \mu)_{T,l_i}$ we can write
\eqa{
\Big (
\frac{\partial }{\partial l_i}
\Big(
\frac{\partial \Omega}{\partial \mu}
\Big )_{T,l_k}
\Big)_{T,l_{k\ne i},N_e}
&\!=&\!0\!=\!\!
\Big(
\frac{\partial ^2 \Omega}{\partial \mu\partial l_i}
\Big)_{T, l_{k\ne i}}\nonumber\\
&\!\!+&\!\! \Big(
\frac{\partial^ 2 \Omega}{\partial^2 \mu}
\Big)_{\! T,l_{k}}
\!\!\Big (
\frac{\partial \mu}{\partial l_i}
\Big)_{\!T,l_{k \ne i},N_e}\label{eq_ddlidomegadmu}\\
\Big (
\frac{\partial }{\partial T}
\Big(
\frac{\partial \Omega}{\partial \mu}
\Big )_{T,l_k}
\Big)_{l_{k},N_e}
&\!=&\!0\!=\!\!
\Big(
\frac{\partial ^2 \Omega}{\partial \mu\partial T}
\Big)_{l_{k}}\nonumber\\
&\!\!+&\!\! \Big(
\frac{\partial^ 2 \Omega}{\partial^2 \mu}
\Big)_{\! T,l_{k}}
\!\!\Big (
\frac{\partial \mu}{\partial T}
\Big)_{\!l_{k},N_e} \label{eq_ddtdomegadmu}
}
Using Eqs. \eqref{eq_ddlidomegadmu} and \eqref{eq_ddtdomegadmu} in
Eqs. \eqref{eq_ddtdomegadli} and \eqref{eq_dsdli} makes it clear that
both expressions [Eqs. \eqref{eq_ddtdomegadli} and \eqref{eq_dsdli}]
are equal and therefore $\alpha_i$ in Eq. \eqref{eq_alphaieqsumij} can be
connected to the derivative of entropy.

\section{Symmetry of $C_{ij}$}
\label{sec_symmetry_cij}
By symmetry $C_{ij}$ should equal $C_{ji}$, which is not directly seen
from Eq. \eqref{eq_cij} since for example $i$-derivative of $\Omega$ is taken at
fixed $\mu$, while $j$-derivative is taken at fixed $N_e$. We show
here for example, that $C_{xy}$  
given with Eq. \eqref{eq_cij} obeys the
symmetry. Keeping in mind that $\Omega=\Omega(T,l_i,\mu)$ and for
fixed $N_e$, $\mu=\mu(T,l_i,N_e)$ we can write out the first term
\eqa{
\Big (
\frac{\partial}{\partial l_y}
\Big(
\frac{\partial \Omega}{\partial l_x}
&&\Big )_{T,l_{k\ne x},\mu}
\Big )_{T,l_{k\ne y},N_e}
=
\Big(
\frac{\partial^2 \Omega}{\partial l_y \partial l_x}
\Big )_{T,l_{k\ne x,y},\mu}\nonumber\\
&&+
\Big(
\frac{\partial^2 \Omega}{\partial l_x \partial \mu}
\Big )_{T,l_{k\ne x}}
\Big(
\frac{\partial \mu}{\partial l_y }
\Big )_{T,l_{k\ne y},N_e}.
\label{eq_ddlydomegadlx}
}
By using $-N_e=(\partial \Omega/\partial \mu)_{T,l_i}$ one obtaines
\eqa{
\Big(
\frac{\partial }{\partial l_x}
\Big(
\frac{\partial \Omega}{\partial \mu}
\Big )_{T,l_i}
\Big)_{T,l_{k\ne x},N_e}
=
0
=
\Big(
\frac{\partial^2 \Omega}{\partial l_x \partial \mu}
\Big)_{T,l_{k\ne x}}\nonumber\\
+
\Big(
\frac{\partial^2\Omega}{\partial \mu^2}
\Big)_{T,l_i}
\Big(
\frac{\partial \mu}{\partial l_x}
\Big)_{T,l_{k\ne x},N_e}.
}
Using  this relation in Eq. \eqref{eq_ddlydomegadlx} and then further
in Eq. \eqref{eq_cij} one gets
\eqa{
C_{xy}&=&
\frac{l_x l_y}{V} 
\Big [
\Big(
\frac{\partial^2 \Omega}{\partial l_x\partial l_y}
\Big )_{T,l_{k\ne x,y},\mu}
\\ &-&
\Big(
\frac{\partial^2\Omega}{\partial \mu^2}
\Big)_{T,l_i}
\Big(
\frac{\partial \mu}{\partial l_x}
\Big)_{T,l_{k\ne x},N_e}
\Big(
\frac{\partial \mu}{\partial l_y}
\Big)_{T,l_{k\ne y},N_e}
\Big]. \nonumber
}
From this it is obvious that $C_{xy}=C_{yx}$ and the symmetry is
obeyed.

\bibliographystyle{apsrev4-1}
\bibliography{ref_wr_texpan}

\end{document}